\documentclass{aa}  

\usepackage{graphicx}
\usepackage[varg]{txfonts}
\bibpunct{(}{)}{;}{a}{}{,}

\newcommand{\oo}{HR\,8799}
\newcommand{\cd}{\,d$^{-1}$}
\newcommand{\most}{{\it MOST}}
\newcommand{\gdor}{$\gamma$~Dor}
\newcommand{\m}{\hphantom{$-$}}
\newcommand{\vsini}{$v \sin i$}

\begin{document} 
\title{
    $MOST$\ light-curve analysis of the $\gamma$~Dor pulsator HR\,8799, showing resonances and amplitude variations\thanks{
    Based on data from the \most\ satellite, a Canadian Space Agency mission, jointly operated by Dynacon Inc., the University of Toronto Institute for Aerospace Studies and the University of British Columbia, with the assistance of the University of Vienna.
  }
}

\author{
  \'A. S\'odor
  \inst{1,2}
  \and
  A.-N. Chen\'e
  \inst{3}
  \and
  P. De Cat
  \inst{1}
  \and
  Zs. Bogn\'ar
  \inst{2}
  \thanks{Currently a voluntary collaborator at the Royal Observatory of Belgium, Ringlaan 3, B-1180 Brussel, Belgium}
  \and
  D. J. Wright
  \inst{1,4}
  \and
  C. Marois
  \inst{5}
  \and
  G.~A.~H. Walker
  \inst{6}
  \and
  J. M. Matthews
  \inst{7}
  \and
  T. Kallinger
  \inst{8}
  \and
  J. F. Rowe
  \inst{9}
  \and
  R. Kuschnig
  \inst{8}
  \and
  D.~B. Guenther
  \inst{10}
  \and
  A.~F.~J. Moffat
  \inst{11}
  \and
  S.~M. Rucinski
  \inst{12}
  \and
  D. Sasselov
  \inst{13}
  \and
  W.~W. Weiss
  \inst{8}
}

\institute{
  Royal Observatory of Belgium, Ringlaan 3, B-1180 Brussel, Belgium;
  \email{adam.sodor@oma.be}
  \and
  Konkoly Observatory, MTA CSFK, Konkoly Thege M. u. 15--17, H--1121 Budapest, Hungary;
  \email{sodor@konkoly.hu}
  \and
  Gemini Observatory, Northern Operations Center, 670 North A'ohoku Place, Hilo, HI, 96720, USA
  \and
  Department of Astrophysics and Optics, School of Physics, University of New South Wales, Sydney 2052, Australia
  \and
  National Research Council Canada, Herzberg Institute of Astrophysics, 5071 West Saanich Road, Victoria, BC V9E 2E7, Canada
  \and
  1234 Hewlett Place, Victoria, BC V8S 4P7, Canada
  \and
  Department of Physics and Astronomy, University of British Columbia, 6224 Agricultural Road, Vancouver, BC V6T 1Z1, Canada
  \and
  Institute for Astronomy (IfA), University of Vienna, T\"urkenschanzstrasse 17, 1180 Vienna, Austria
  \and
  NASA Ames Research Park, MS 244-30, Building 244, Room 107A, Moffett Field, CA 94035-1000
  \and
  Institute for Computational Astrophysics, Department of Astronomy and Physics, Saint Marys University, Halifax, NS B3H 3C3, Canada
  \and
  D\'epartement de physique, Universit\'e de Montr\'eal, C.P. 6128, Succ. Centre-Ville, Montr\'eal, QC H3C 3J7, and Observatoire du mont M\'egantic, Canada
  \and
  Department of Astronomy and Astrophysics, David Dunlap Observatory, University of Toronto, P.O. Box 360, Richmond Hill, ON L4C 4Y6, Canada
  \and
  Harvard--Smithsonian Center for Astrophysics, 60 Garden Street, Cambridge, MA 02138
}

\date{Received  ; accepted }

 
\abstract
 {The central star of the \object{HR\,8799} system is a $\gamma$~Doradus-type pulsator. The system harbours four planetary-mass companions detected by direct imaging, and is a good solar system analogue. The masses of the companions are not known accurately, because the estimation depends strongly on the age of the system, which is also not known with sufficient accuracy. Asteroseismic studies of the star might help to better constrain the age of HR\,8799. We organized an extensive photometric and multi-site spectroscopic observing campaign for studying the pulsations of the central star.}
 {The aim of the present study is to investigate the pulsation properties of HR\,8799 in detail via the ultra-precise 47-d-long nearly continuous photometry obtained with the {\it MOST} space telescope, and to find as many independent pulsation modes as possible, which is the prerequisite of an asteroseismic age determination.}
 {We carried out Fourier analysis of the wide-band photometric time series.}
 {We find that resonance and sudden amplitude changes characterize the pulsation of HR\,8799. The dominant frequency is always at \hbox{$f_1 = 1.978\,\mathrm{d}^{-1}$.} Many multiples of one ninth of the dominant frequency appear in the Fourier spectrum of the {\it MOST} data: $n/9\,f_1$, where $n=\{1, 2, 3, 4, 5, 6, 7, 8, 9, 10, 13, 14, 17, 18\}$. Our analysis also reveals that many of these peaks show strong amplitude decrease and phase variations even on the 47-d time-scale. The dependencies between the pulsation frequencies of HR\,8799 make the planned subsequent asteroseismic analysis rather difficult. We point out some resemblance between the light curve of HR\,8799 and the modulated pulsation light curves of Blazhko RR~Lyrae stars.}
 {}


\maketitle

\begin{figure*}[!!!hhhhttttt]
  \begin{center}
  \includegraphics[width=14.5cm]{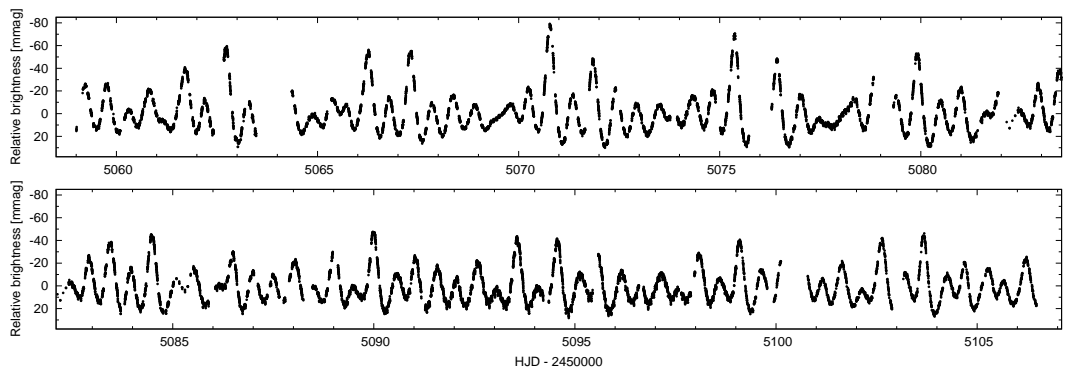}
  \end{center}
  \caption{The cleaned \most\ light curve of \oo, containing 57\,094 data points. The visible periodic gaps are due to the removal of the data points affected by stray light contamination reflected from the surface of the Earth.}
  \label{fig:mostlc}
\end{figure*}

\section{Introduction}

\oo\ (HD\,218396, V342~Peg, $\alpha_{2000} = 23^{\rm h}07^{\rm m}28.\!\!^{\rm s}71$, $\delta_{2000} = +21{\degr}08'03.\!\!^{\prime\prime}3$) is a bright ($V=5.95$\,mag) A5V star, showing $\lambda$-Bootis-type metallic-abundance anomalies \citep{Sadakane2006PASJ...58.1023S}, and \gdor\ pulsations \citep{Zerbi1999MNRAS.303..275Z}.

The system of \oo\ is one of the very few known true solar system analogues \citep{MatthewsB2014ApJ...780...97M}. It hosts four planetary-mass companions detected by direct imaging \citep{Marois2008Sci...322.1348M,Marois2010Natur.468.1080M}, and circumstellar dust disk with warm, cold and external halo components \citep{Sadakane1986PASP...98..685S,Reidemeister2009A&A...503..247R,Su2009ApJ...705..314S}. The debate over the age of the system is still unsettled, as different methods give significantly different ages \citep[see, e.g.,][]{MoroMartin2010ApJ...721L.199M}. Since the masses of the companions are derived from their luminosities using theoretical cooling tracks, the mass determination is very sensitive to the age of the system. 

The oscillations of \oo\ were previously studied via ground-based photometric campaigns by \citet[][Z99 hereafter]{Zerbi1999MNRAS.303..275Z} and \citet[][C09 hereafter]{Cuypers2009A&A...499..967C}, and spectroscopically by \cite{Mathias2004A&A...417..189M} and \cite{Wright2011ApJ...728L..20W}.

The first asteroseismic age determination was performed by \cite{Moya2010MNRAS.405L..81M}, who derived different possible age ranges based on three pulsation frequencies published by Z99, using the frequency ratio method. The obtained asteroseismic age is not well constrained; it depends on the inclination of the star. If the inclination is either around 36$\degr$ or around 50$\degr$, then the possible age ranges are either [26, 430] and [1123, 1625], or [1126, 1486] Gyr, respectively. The inclination of the rotation axis of the central star is often assumed to be aligned with the inclination of the planetary system. On one hand, the orbits of the companions are seen about face-on, but the real inclination angles are not well constrained, because the orbital periods are in the 50--500\,yr range, and the orbital displacements of the companions were small since their first detection. The best estimation of the inclination of the orbital plane, assuming co-planarity between the companions, is $i_\mathrm{orb} = 28\degr$ \citep{Esposito2013A&A...549A..52E}. On the other hand, analysis of the spectroscopic line-profile variations, caused by the pulsation, suggests that the central star is visible at higher inclination; $i_\mathrm{rot} \gtrsim 40\degr $ \citep{Wright2011ApJ...728L..20W}. Note that alignment between the rotation and pulsation axis is assumed in this estimation. Thus, a misalignment of $\Delta i > 10\degr$ between the inclination of the orbits and the rotational inclination of the central star seems probable.

To facilitate the asteroseismic study of \oo, we obtained photometric observations with the \most\ space telescope in 2009, and also organized an extended multi-site spectroscopic observing campaign around the time of the \most\ observations. The light curve shows interesting phenomena detected at the first time in \gdor\ pulsators. The interpretation of the spectroscopic data is a complex task in itself, therefore, we analyse only the photometric data in the present study. The investigation of the spectroscopic time series is planned in a forthcoming paper.

\section[]{Observations and data preparation}

This study is based on the space photometry obtained by the Microvariability and Oscillations of STars (\most) microsatellite nearly continuously over a 47-d-long period in 2009, between 15 August and 1 October (HJD\,2\,455\,059--106).

\most\ contains a 15-cm Rumak-Maksutov telescope imaging onto a CCD detector via a custom optical broadband filter (350--750\,nm; \citealt{Walker2003PASP..115.1023W, Matthews2004Natur.430...51M}). From its polar Sun-synchronous orbit of altitude 820\,km and period 101 min, \most\ has a continuous viewing zone of
about 54$^\circ$ wide within which it can monitor target fields for up to two months without interruption. Targets as bright as \oo\ are projected onto the science CCD as a fixed Fabry image of the telescope entrance pupil covering some 15\,000 pixels. The photometry is non-differential, but given the orbit, thermal and design characteristics of \most, experience has shown that it is a very photometrically stable platform even over long timescales (with repeatability of the mean instrumental flux from a non-variable target of the brightness of \oo\ to within about 1 mmag over a month). 

\begin{figure*}
  \begin{center}
  \includegraphics[width=14.6cm]{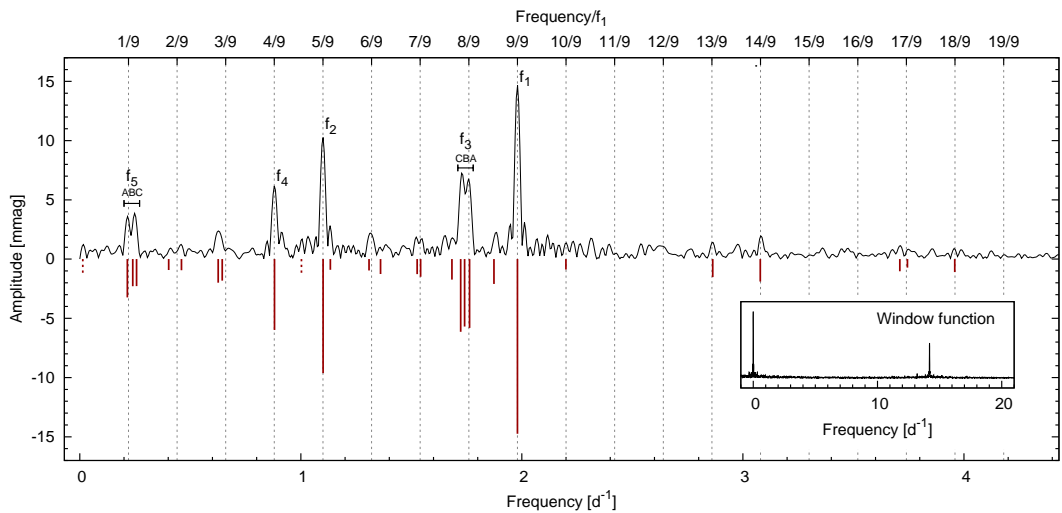}
  \end{center}
  \caption{Fourier amplitude spectrum of the \most\ light curve of \oo. The actual spectrum is shown in the positive amplitude range, while the 28-frequency solution, listed in Table~\ref{tbl:fullfreq}, is represented by thick vertical lines in the negative amplitude range. Instrumental frequency components are marked by dotted lines. The most prominent frequencies and frequency groups are labelled. The $n/9\,f_1$ frequency grid is indicated by dashed gray lines. The insert shows the window function of the data. The only high-amplitude alias is at the orbital frequency of the satellite, which causes aliasing only far out of the investigated range.}
  \label{fig:fullsp}
\end{figure*}

Due to the failure of the tracking CCD, the science exposures on the \most\ science CCD now take place at the same time as the guide-star exposures for satellite pointing. In the case of the \oo\ observations, the guide-star exposure time (and hence also the science target exposure time) was 0.3~sec. To build up sufficient signal-to-noise ratio (S/N), the science exposures were added on board the satellite in ``stacks'' of 100 exposures, each stack corresponding to 30~sec of total integration. Stacks were downloaded from the satellite consecutively, with no dead time between stacks, giving a sampling rate of about twice per minute. A total of 101,954 individual data-points have been obtained. However, due to a very high background occurring at each orbit caused by the stray light contamination reflected from the surface of the Earth, a total of $\sim$40\,000 points had to be removed, leaving periodic gaps in the light curve.\footnote{In fact, a 0.1 phase range is completely missing from the light curve, if phased with the orbital frequency, 14.1883\cd. This causes the prominent peak at this frequency in the spectral window function, shown in Fig.~\ref{fig:fullsp}} After removing several other outliers, we obtained the final cleaned light curve consisting in 57\,094 data points, shown in Fig.~\ref{fig:mostlc}. We used these data for analysis. The typical $1\sigma$ errorbar of a single data point is 1\,mmag. The light curve is available from the authors upon request.

\section{Light-curve analysis}

We used the {\sc LCfit} \citep{lcfit} software and other in-house developed tools for the following analysis. The frequency identification and light-curve prewhitening was performed in a standard way. A more detailed description of the procedure can be found in \cite{Sodor2014MNRAS.438.3535S}.

\subsection{Frequency identification}

The Fourier amplitude spectrum of the light curve is shown in Fig.~\ref{fig:fullsp}. We looked for significant periodicities with $\mathrm{S/N} > 4$ in the 0--10\cd\ frequency range. We do not investigate higher frequency domains, because those parts of the spectrum are contaminated by artefacts originating from the satellite and caused by the 14.2\cd\ aliasing of the orbital frequency. Since \gdor\ pulsation frequencies are below 10\cd, this limitation does not affect our study.

Altogether, 28 significant frequency components were identified with $\mathrm{S/N} > 4$. The noise level was calculated before each prewhitening step as a moving average of the amplitude spectrum with a 5\cd\ window size. Two of the significant frequencies are identified as instrumental artefacts. \hbox{$f_\mathrm{sid}=1.0026$\cd} is the sidereal frequency, while the lowest-frequency component, \hbox{$f_\mathrm{low}=0.013$\cd,} which is below the Rayleigh limit\footnote{The Rayleigh resolution limit is calculated as $1/T$, where $T$ is the total time span of the light curve.}, thus describes a periodicity longer than the observation period, is also not considered to be intrinsic to the star. The frequencies and amplitudes of the 28-frequency solution are plotted together with the Fourier periodogram in Fig.~\ref{fig:fullsp}, and are listed in Table~\ref{tbl:fullfreq}. Note that the stellar-origin and instrumental frequencies were treated in the same way during prewhitening. 

\begin{table*}
  \caption{Frequencies detected with $\mathrm{S/N} > 4$ in the \most\ light curve of \oo. Standard errors of the frequencies are given in parentheses in the unit of the last digit. Cols. 6 and 8 give the deviation of the actual frequency (Col. 2) from the linear combination and from the exact $n/9\,f_1$ resonance (Cols. 5 and 7), respectively, and are given in the unit of the Rayleigh resolution limit ($Rl = 0.021$\cd). The frequencies in the bottom two lines are not considered to be intrinsic to \oo.}
  \label{tbl:fullfreq}
  \centering
  \begin{tabular}{lrrrclrl}
    \hline
    ID               & Frequency   & Ampl.& S/N  &linear combination& deviation &  resonance  & deviation\\
                     & (d$^{-1}$)  &(mmag)&      &                  & ($Rl$)    & ($1/9\,f_1$)& ($Rl$)\\
    \hline
    $f_{1}$          & 1.97984(01) & 14.74& 70.6 & \\
    $f_{2}$          & 1.10033(02) & 9.62 & 42.9 &                  &           &  5 & \m0.02 \\
    $f_{3\mathrm A}$ & 1.76316(09) & 5.81 & 25.4 & $2 (f_1-f_2)$    &  \m0.2    &  8 & \m0.2  \\
    $f_{3\mathrm B}$ & 1.74065(11) & 5.69 & 25.7 & $2 (f_1-f_2)$    & $-0.9$    &  8 & $-0.9$ \\
    $f_{3\mathrm C}$ & 1.72305(13) & 6.14 & 31.8 & $2 (f_1-f_2)$    &  $-1.7$   &  8 & $-1.8$ \\
    $f_{4}$          & 0.88071(04) & 5.98 & 29.4 & $f_1 - f_2$      &  \m0.06   &  4 & \m0.04 \\
    $f_{5\mathrm A}$ & 0.21522(12) & 3.24 & 14.4 & $f_1 - f_{3A}$   &  $-0.1$   &  1 & $-0.2$ \\
    $f_{5\mathrm B}$ & 0.23959(23) & 2.29 & 10.2 & $f_1 - f_{3B}$   &  \m0.002  &  1 & \m1.0  \\
    $f_{5\mathrm C}$ & 0.25646(25) & 2.28 & 9.4  & $f_1 - f_{3C}$   &  $-0.01$  &  1 & \m1.7  \\
    $f_{6}$          & 3.07858(12) & 1.90 & 10.6 & $f_1 + f_2$      &  $-0.08$  & 14 & $-0.05$\\
    $f_{7}$          & 1.87386(12) & 2.10 & 9.9  & \\
    $f_{8\mathrm B}$ & 0.64421(18) & 1.81 & 8.0  & $f_{3B} - f_2$   &  \m0.2    &  3 & $-0.7$ \\
    $f_{8\mathrm C}$ & 0.62575(17) & 1.99 & 8.9  & $f_{3C} - f_2$   &  \m0.14   &  3 & $-1.6$ \\
    $f_{9}$          & 1.68339(19) & 1.73 & 8.4  & \\
    $f_{10}$         & 2.86332(15) & 1.52 & 8.2  & $2f_1-f_2$       &  \m0.2    & 13 & \m0.2  \\
    $f_{11}$         & 3.95842(21) & 1.10 & 6.8  & $2f_1$           &  $-0.06$  & 18 & $-0.06$\\
    $f_{12\mathrm A}$& 3.74450(35) & 0.72 & 4.5  & $f_1+f_{3A}$     &  \m0.3    & 17 & \m0.2  \\
    $f_{12\mathrm C}$& 3.71023(24) & 1.03 & 6.3  & $f_1+f_{3C}$     &  \m0.3    & 17 & $-1.4$ \\
    $f_{13\mathrm A}$& 1.54171(30) & 1.49 & 6.2  &                  &           &  7 & \m0.1  \\
    $f_{13\mathrm B}$& 1.52643(36) & 1.27 & 5.3  &                  &           &  7 & $-0.6$ \\
    $f_{14\mathrm A}$& 1.30868(26) & 0.96 & 4.7  & $f_1+f_2-f_{3A}$ &  $-0.4$   &  6 & $-0.5$ \\
    $f_{14\mathrm C}$& 1.36057(20) & 1.26 & 5.7  & $f_1+f_2-f_{3C}$ &  \m0.2    &  6 & \m1.9  \\
    $f_{15\mathrm A}$& 0.45956(26) & 0.94 & 4.5  &                  &           &  2 & \m0.9  \\
    $f_{15\mathrm B}$& 0.40207(27) & 0.92 & 4.2  &                  &           &  2 & $-1.8$ \\
    $f_{16}$         & 1.13345(31) & 0.91 & 4.3  &                  &           &  5 & \m1.6  \\
    $f_{17}$         & 2.19911(27) & 0.89 & 4.0  & $2f_2$           & $-0.1$    & 10 & $-0.03$\\[2mm]
    $f_\mathrm{sid}$ & 1.00261(19) & 1.30 & 5.4  & \\
    $f_\mathrm{low}$ & 0.01344(38) & 1.43 & 5.8  & \\
    \hline
  \end{tabular}
\end{table*}

\begin{figure}
  \begin{center}
  \includegraphics[width=8cm]{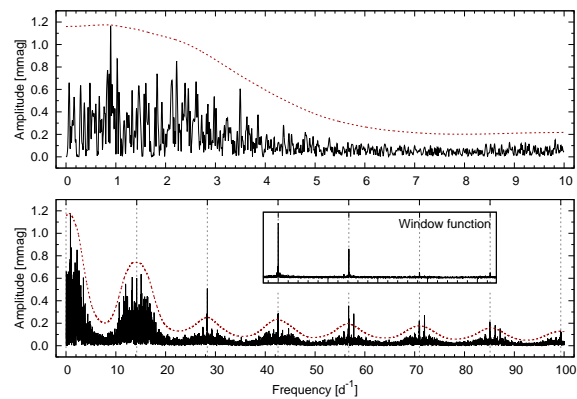}
  \end{center}
  \caption{Prewhitened Fourier spectrum of the \most\ light curve of \oo, after the subtraction of the 28 frequencies listed in Table~\ref{tbl:fullfreq}. Dashed curves show the $4\sigma$ noise level. Vertical dashed lines in the bottom panel and in the 'Window function' insert panel mark the multiples of the orbital frequency ($f_\mathrm{orb}=14.2$\cd) of \most. The bottom panel shows the residual spectrum up to 100\cd, while the top panel is a blow-up of the investigated frequency range.}
  \label{fig:ressp}
\end{figure}

The $(O-C)$ of the 28-frequency fit is 3.2\,mmag, significantly larger than the 1\,mmag uncertainty of the individual data points. The residual spectrum is plotted in Fig.~\ref{fig:ressp}. It shows extended noise below 5\cd, which is partially due to residual peaks around the detected significant peaks caused by the amplitude and frequency variations (see details in Sect.~\ref{sect:apvar}). Further signals intrinsic to the star but of amplitudes below the detection limit might contribute to the increased noise in this low-frequency range. Also, instrumental noise is typically higher at low frequencies.

The bottom panel of Fig.~\ref{fig:ressp} plots the residual spectrum of the light curve up to 100\cd. We can see an increase in the noise level around each harmonic of the orbital frequency, $f_\mathrm{orb}=14.2$\cd, which is the aliasing of the extended noise below 5\cd. Only several harmonics of $f_\mathrm{orb}$ and high-order linear combinations of $f_\mathrm{orb}$ and $f_\mathrm{sid}$ exceed the $4\sigma$ noise level.

There is no sign of $\delta$~Scuti pulsation with amplitude higher than $\sim0.7$\,mmag in the 10--20\cd\ range and with amplitude higher than $\sim0.2$\,mmag above 20\cd\ in \oo.

\subsection{Resonances}

The light variations and the Fourier periodogram are dominated by a strong signal at $f_1=1.98$\cd, while the second highest peak is at $f_2=1.10$\cd. The ratio of these two frequencies is $f_1/f_2 = 1.8 = 9/5$, showing resonance. Note that $f_2$ is off only by 0.02-times the Rayleigh resolution from $5/9\,f_1$, that is, the resonance is well established. Furthermore, the arrangement of the significant frequencies follows a conspicuous regular pattern. Many frequencies are aligned with the integer multiples of $f_1/9$. This frequency grid is also indicated in Fig.~\ref{fig:fullsp}.

\subsection{Linear combinations}

In connection with the alignment of the frequencies with the $n/9\,f_1$ grid, many linear-combination frequencies can be identified. However, the parent--child or base--combination relations in the linear-combination identification are ambiguous.

A simple description to explain the frequency alignments with the $n/9\,f_1$ grid would be to choose $f_{5\mathrm A} \approx 1/9\,f_1$ as the base frequency, in which case, the other aligned frequencies would be integer multiples of this one. However, $f_{5\mathrm A}$ itself is not quite well aligned with the $n/9\,f_1$ grid, and such a solution does not take into account the amplitude relations of the components.

A standard method for identifying linear combinations is to demand both base frequencies to have higher amplitudes than that of the combination frequency. Following this way, we can identify five base frequencies ($f_1, f_2, f_{3\mathrm A}, f_{3\mathrm B}, f_{3\mathrm C}$), of which the latter three are related, and many combinations of these. Only two intrinsic frequencies ($f_7$ and $f_9$) cannot be described either as combination of these five bases or $n/9\,f_1$ multiples. This identification is listed in Table~\ref{tbl:fullfreq}.

\subsection{Amplitude and phase variations}
\label{sect:apvar}

The Fourier decomposition of the observed time series assumes stationary harmonic components. If this assumption is not met, in the case when the amplitude and/or phase of a sinusoidal component changes during the observations, the Fourier analysis of the data will show bunching of frequencies in a small range, with separations of the order of the Rayleigh resolution.

We find such close frequencies in \oo, forming groups. We denote the group members with A, B, and C subscripts in Table~\ref{tbl:fullfreq}, while we refer to the whole group by omitting the A, B, and C subscripts. The most prominent of these groups is the $f_3$ group around $8/9\,f_1$, which, together with $f_1$ and $f_2$, appears to be a linear-combination base of other groups. The highest-amplitude linear-combination group is $f_5 = f_1-f_3$, around $1/9\,f_1$.

On one hand, the $f_3$ group and its linear-combination groups might represent amplitude and phase variations of a single pulsation mode and its linear-combinations, respectively. On the other hand, it is also possible that the group components are independent frequencies present in \oo, and we actually observe beating of these poorly resolved components. The limited time coverage of 47\,d of the \most\ dataset makes it difficult to discriminate between these two possibilities.\footnote{Unfortunately, the longer time-base spectroscopic data are also not quite suitable to decide this question. Even though the frequency resolution of the spectroscopic data is much better than that of the presently studied photometry, the uneven distribution of the spectroscopic data poses another difficulty. We will revisit this question in the forthcoming spectroscopy paper on \oo.}

\begin{figure*}
  \begin{center}
  \includegraphics[width=14.6cm]{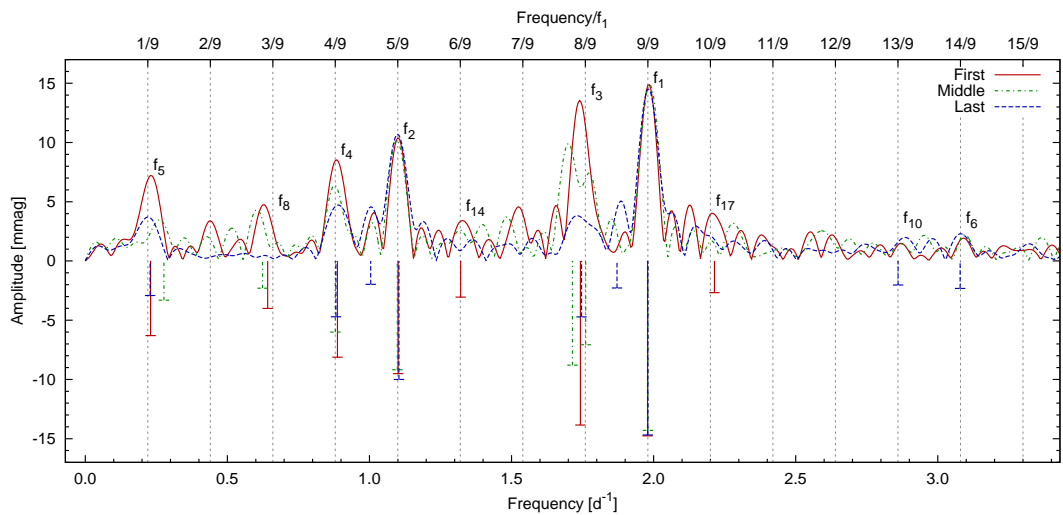}
  \end{center}
  \caption{Variations in the frequency content of the \most\ light curve of \oo. Fourier amplitude spectra of the first (gray/red full line), middle (gray/green dash-dotted line) and last (black/blue dashed line) third of the data are plotted with curves in the positive amplitude range, while the frequency components identified in the respective time segments, down to the $4\sigma$ level, are plotted with vertical lines in the negative amplitude range. The horizontal lines at the lower end of these lines are added only for better visibility of the overlapping ones.}
  \label{fig:freqvar}
\end{figure*}

We investigated the suspected variations in the frequency components by dividing the data into three equally long subsets, each covering 15.5\,d. We performed the same frequency identification on the subsets as previously on the full dataset. The Fourier periodograms and the identified frequency components of the subsets are plotted in Fig.~\ref{fig:freqvar}.

The spectra of the three subsets clearly show variations. Such variations are expected in any of the two possible cases, described above. Only the amplitude of $f_1$ and $f_2$ remained stable during the 47-d-long \most\  observations, while all the other components detected in the shorter subsets underwent amplitude decrease.\footnote{Note that $f_{10}$ and $f_6$ were only detected in the last time segment. These two frequencies have amplitudes around the detection limit, and their detection only in the last third can be explained by the different noise statistics of the three data subsets. The $1\sigma$ levels of the three residual spectra are 0.08, 0.06 and 0.04\,mmag around 9\cd\ for the first, middle and last segments, respectively.} The observed overall amplitude decrease in many frequency components suggests that we indeed see amplitude variations instead of beating between close, poorly resolved frequency components. In the latter case, both amplitude increase and decrease would be expected among the changing components. The double peak of $f_3$ and the frequency shifts of $f_3$, $f_5$ and $f_8$ in the middle time segment hints that most of the variations occurred during this interval.

A single gradual amplitude decrease event does not cause multiple peaks in the resultant Fourier spectrum; only the corresponding peak becomes somewhat broadened. Thus, the peak multiplets that form the groups in the Fourier spectrum of \oo, are signs of frequency or phase variations accompanying the amplitude decrease.

\cite{Breger2014ApJ...783...89B} identified parent--child relations of resonantly coupled linear-combination frequencies in the extended {\it Kepler} light curve of a $\delta$~Scuti star, KIC\,8054146. Their identification scheme relies on relations between the amplitudes and phases of the parents and their child component. Since not all the terms are known in their formula (eq. 5), it is necessary to detect amplitude and/or phase variations for establishing the parent--child relation. Observing variations in both parents, they were able to predict the amplitude and phase of the child with an unexpected accuracy for 20 separate time segments.

Our data on \oo\ are not extended enough to make similar quantitative predictions, but we can check our linear-combination identifications qualitatively, at least for the highest-amplitude combination frequencies, $f_4$, $f_5$ and $f_8$. Among the three frequencies considered to be bases ($f_1$, $f_2$ and $f_3$), only the last one changed. 

Both $f_5$ and $f_8$ show a strong amplitude decrease from the first to the third time segment. In fact, $f_8$ disappeared completely. This, and the apparent frequency shifts and the frequency fine-structure in the complete dataset suggests that $f_5$ and $f_8$ are indeed linear combinations of $f_3$ with $f_1$ and $f_2$, respectively, as Table~\ref{tbl:fullfreq} lists.

We cannot explain, however, the amplitude decrease of $f_4$ this way. The straightforward explanation for the frequency itself would be $f_4 = f_1-f_2$, as listed in Table~\ref{tbl:fullfreq}, but in this case we would not expect amplitude variation, since both parents are stable. We assume that $f_3$ is also involved in this component. An alternative expression is $f_4 = f_2 - (f_1 - f_3) = f_2 - f_1 + f_3$ to describe this component. This is possible because $f_4$ is in 1:2 resonance with $f_3$. $f_3$ can influence the amplitude of $f_4$ either through the linear combination or through the 1:2 resonance.

\section{Blazhko-like behaviour of \oo}

The amplitude and phase modulation of the light curves observed in a large fraction of RR~Lyrae stars, the so-called Blazhko effect, is a long-standing, still unresolved problem of astrophysics \citep{Kolenberg2012JAVSO..40..481K}. Recent continuous observations of the {\it Kepler} space telescope revealed that period doubling (PD) -- alternating higher and lower light maxima and minima -- also often occur in Blazhko RR~Lyrae stars \citep{Kolenberg2010ApJ...713L.198K}.

Although the physical background of the phenomenon might be completely different in \oo, we point out some interesting resemblance of the light variations and the Fourier spectra of this star and those of Blazhko RR~Lyrae stars.

\begin{figure}
  \begin{center}
  \includegraphics[width=7cm]{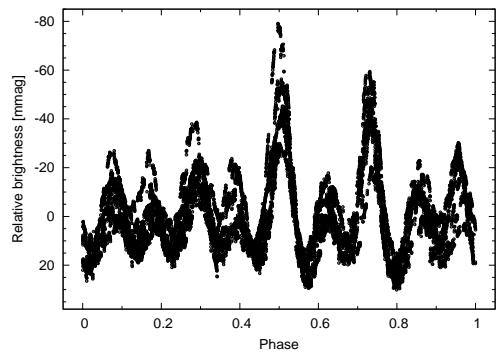}
  \end{center}
  \caption{The light curve of \oo\ phased with the beating frequency $1/9\,f_1 \approx f_1-f_{3\mathrm A}$. The time zero point was chosen to give the main peak at phase 0.5.}
  \label{fig:lcfold}
\end{figure}

A periodic amplitude modulation of the pulsation can be seen in Fig.~\ref{fig:mostlc}. The modulation period of about 4.5\,d corresponds to $1/9\,f_1 \approx f_1-f_{3\mathrm A}$, that is, we see beating between two strong periodicities. The light curve phased with $1/9\,f_1$ is shown in Fig.~\ref{fig:lcfold}. This plot shows not only the nice periodic amplitude modulation, but also PD: the brightest maxima at phase 0.5 are always preceded by weak ones, and are followed by weak ones, then stronger ones, then weaker maxima again.

Amplitude and phase modulation appear in the frequency domain as modulation side-peaks situated around a main peak with equidistant separation  \citep{Szeidl2009CoAst.160...17S}. PD corresponds to half-integer frequencies of the main peak in frequency domain. Z99 already raised the possibility that we actually see modulation of the dominant periodicity of \oo\ in the periodogram, as the dominant frequency, $f_1$, was accompanied by another one at 1.73\cd\ in their data ($f_3$ in our analysis). However, they rejected this explanation afterwards, as they could not find the higher-frequency modulation side peak around $f_1$, with amplitude approximately equal to the lower-frequency one. Later, theoretical studies and observations showed that the amplitudes of the lower- and higher-frequency modulation components may differ significantly \citep{Szeidl2009CoAst.160...17S}, even to the extent that the lower-amplitude one remains undetected \citep{Jurcsik2005AcA....55..303J}.

All these provide the basis for an alternative explanation of the structure of the Fourier spectrum of \oo. According to this description, $f_1$ and $f_2$ are still base frequencies, while the third one is $f_\mathrm{mod} = 1/9\,f_1 \approx f_5$, the modulation frequency. In Fig.~\ref{fig:fullsp}, we can identify the symmetrically situated modulation peaks ($\pm f_\mathrm{mod}$) around $f_1$, $f_2$, $2f_1$, and $f_1+f_2$. The lower-frequency side-peaks have always much higher amplitudes than the higher-frequency ones, so that these are detectable above the noise only for the two highest-amplitude peaks, $f_1$ and $f_2$. The \most\ data actually reveal the higher-frequency modulation side peak of the dominant pulsation mode: $f_1+f_\mathrm{mod} = f_{17}$, which could not be detected by Z99. The modulation frequency in this description undergoes variations even on the 47-d time span of the data. Although on longer timescales, such variations are known in Blazhko RR~Lyrae stars, too \citep[e.g.,][]{Sodor2011MNRAS.411.1585S}.

Another possibility is that $f_2$ is, in fact, not an independent pulsation mode, but, together with $f_4$, is the expression of PD. The PD is related to the 9:5 and 9:4 ratio between $f_1$:$f_2$ and $f_1$:$f_4$, respectively. As often is the case also in RR~Lyrae stars, the half-integer frequencies are offset somewhat from the exact $(2k+1)/2\, f_\mathrm{puls}$ values, in relation with the temporal variation in the PD itself \citep{Szabo2010MNRAS.409.1244S}. The large offset of $\pm 1/18\,f_1$ in \oo\ is necessary because of the resonances involving the number nine, which is not divisible by two, and because of the quite regular nature of the PD. Due to the 1:9 resonance, the length of one modulation cycle is exactly nine times the length of the dominant pulsation cycle. Furthermore, as Fig.~\ref{fig:lcfold} demonstrates, a high maximum occurs for every ninth pulsation. Therefore, the high-low-high amplitude alternation has to turn to the other way exactly within one modulation cycle, and then back to the original within another one. All these can be considered as the $1/2\,f_1$ and $3/2\,f_1$ half-integer frequencies of the PD are modulated with the $f_\mathrm{mod}/2$ frequency. We indeed see $f_2$ and $f_4$ exactly at $1/2\,f_1 \pm f_\mathrm{mod}/2$, while, $f_{10}$ and $f_6$ are situated exactly at $3/2\,f_1 \pm f_\mathrm{mod}/2$.

\section{Discussion}

The pulsation of \oo, as revealed by the ultra-precise \most\ photometry, is not a typical multi-periodic \gdor\ pulsation. We detected significant amplitude and frequency variations even during the 47 days of the \most\ observations. Resonance also plays a very important role in the pulsation of \oo. Two previously published studies already investigated the frequency content of \oo\ in photometric observations (Z99,C09); therefore, we discuss our frequency analysis results in comparison with the earlier analyses.

\subsection{$f_1 = 1.98$\cd}

The only frequency that can be considered with certainty to correspond to an independent mode is the dominant one, $f_1$. This signal was detected in all previous photometric runs, with frequencies agreeing within the corresponding Rayleigh resolution limits. The measured amplitudes are also similar in each epoch. This frequency was always the dominant one.

Two earlier works published frequencies and corresponding multi-colour amplitudes from time-series photometric runs of \oo. Since these works are based on different photometric bands, direct comparison of the amplitudes is not possible, but we still can perform a rough quantitative analysis. Z99 (table 2) give a Johnson $V$ amplitude of 16\,mmag for this frequency. C09 (table 24) list amplitudes in seven Geneva bands, where $V$ is 16\,mmag, too, while the highest amplitude, 22.2 mmag is seen in the $B2$ band, corresponding to 450\,nm wavelength. The white-light passband of \most\ covers the 350--750\,nm wavelength range, including the whole Geneva range and extending further towards the red. The 14.7\,mmag amplitude of $f_1$ in the \most\ data is consistent with the previous observations because of the large contribution of red light where the pulsation amplitude is lower. Consequently, there is no sign of significant frequency and amplitude variation of the $f_1$ component over the 14-yr time base covered by the three datasets (between 1995 and 2009).

\subsection{$f_2 = 1.10$\cd}

This component is stable over the time span of the the \most\ observations, and it is also the second strongest signal in the full dataset. Nevertheless, it was not reported by earlier studies. This means that this frequency either was not present in earlier data at all, or its amplitude was much lower than during the \most\ measurements. Checking the residual spectrum of the data of Z99 in the lower left panel of their fig. 2, we can estimate the upper limit of the amplitude of this frequency at that time. Its amplitude was no more than $1/5$ of the amplitude of $f_1$.

This frequency is in 5:9 resonance with the dominant frequency, $f_1$, which might be responsible for its temporal excitation.

Another possibility is that this frequency corresponds to the rotation of the star, if we assume a temperature spot on the surface that modulates the observed brightness. A star spot with life time and migration time scale longer than the 47-d-long \most\ data, but shorter than several years explains the stability of this periodicity over our observing run, and also why this frequency was not detected in earlier data. Using stellar parameters of \citet[][table 2]{Wright2011ApJ...728L..20W}, this rotation frequency yields an inclination in the 23--37$\degr$ range, similar to the orbital inclination.

Note that we detected a low-amplitude peak, $f_{16}$, separated only by 1.6 times the Rayleigh resolution limit from $f_2$. This component might be an indication of a slight amplitude and/or phase change in $f_2$. However, according to Fig.~\ref{fig:freqvar}, its effect on $f_2$ is marginal during the \most\ observing run.

\subsection{$f_3 \approx 1.75$\cd\ and $f_1-f_3 \approx 0.23$\cd}

The signal we denote as $f_3$ appears in each of the three photometric investigations with relatively high amplitude. Its amplitude is only slightly below that of the dominant $f_1$ in the Geneva photometry of C09, while Z99 report its amplitude being only about half of the amplitude of $f_1$. The difference of this and the dominant frequency, $f_1 - f_3 \approx 0.23$\cd\ (or its $-1$\cd\ alias, 0.77\cd) is also detected by all three studies.

This component is highly variable in the \most\ data, both in amplitude and in frequency, while there is no sign of frequency variation in the 4-season-long Geneva photometry of C09. Also, the 1.73\cd\ frequency found by C09 is consistent with that of Z99.

According to Fig.~\ref{fig:freqvar}, $f_3$ is almost in 8:9 resonance with $f_1$, however its actual value is apparently below the exact resonance. This slight off-resonance with the dominant frequency might cause the rather short-term amplitude and frequency changes. Such behaviour can be induced as the assumed resonantly coupled mode is sometimes phase-locked with the exact resonance frequency, while at other times it is not. The somewhat lower frequency of this signal reported by Z99 and C09 was probably sufficiently off-resonance at those times, thus the amplitude was probably not influenced by resonance with $f_1$.

\subsection{$f_9 = 1.68$\cd}

We note the possibility that the third frequency component Z99 found at 1.65\cd\ with amplitude similar to that of the component at 1.73\cd, might originate from frequency and amplitude variations of one signal. On the other hand, it also might be an independent frequency. We also identified an apparently independent frequency component, $f_9 = 1.68$\cd, marginally consistent with the third frequency of Z99. Regardless that it corresponds to the same pulsation mode as $f_9$ of the \most\ data or not, the 1.65\cd\ component of Z99 also underwent amplitude changes. It was undetected in the C09 data, while its amplitude decreased significantly for the \most\ observations.

\subsection{$n$:9 resonances}

The earlier studies of \oo\ did not find resonances between the detected frequencies.

The $n$:9 resonances between most of the identified frequencies and the dominant frequency in \oo\ is a new phenomenon amongst \gdor\ pulsators. No similar $n$:$m$ resonances were reported in any \gdor\ variable previously. Also, only one $\delta$~Sct pulsator is known where $n$:3 resonances were identified, quite recently, in CoRoT space photometry (HD\,51844, Hareter et al., submitted for publication).

The fact that all the identified resonances are of $n/9\,f_1$ form, and that no other than $1/9$ fractions of the dominant frequency are involved, suggest that the $1/9\,f_1$ frequency itself has some physical relevance. However, the $1/9\,f_1$ component, $f_5$, seems to be much more unstable, both in amplitude, and in frequency or phase, than $f_1$ itself. We note that $1/9\,f_1 = 0.22$\cd\ definitely cannot be the rotation frequency of \oo. Considering the stellar parameters given by \cite{Wright2011ApJ...728L..20W}, calculating with the most extreme parameter values from the uncertainty ranges ($R=1.8\,R_\odot$, \vsini\,$=36$\,km/s, $i=90\degr$), the lowest possible rotation frequency is 0.39\cd.

The physical explanation of the $n$:9 resonances is unknown, and they add another puzzle to asteroseismology.

\section{Summary}

\begin{itemize}
 \item 47 days of almost continuous light curve of \oo\ obtained by the \most\ spacecraft was analysed.
 \item 26 intrinsic frequencies were identified.
 \item Many of these frequencies are aligned with an $n/9\,f_1$ grid defined by the dominant frequency, $f_1=1.98$\cd. This alignment suggests resonant coupling between the aligned frequencies and $f_1$.
 \item Some frequencies form groups, suggesting a single underlying signal that underwent amplitude decrease and frequency variations during the 47-d-long observing run.
 \item The light curve of \oo\ and its Fourier periodogram shows interesting resemblance to the light curves and Fourier spectra of Blazhko RR~Lyrae pulsators. The modulation frequency is $\sim1/9\,f_1$, and shows short-term variations. Strong period doubling is also visible.
\end{itemize}

\section{Conclusion}

The identification of independent pulsation modes in the \most\ light curve and in previously published photometric data of \oo\ is rather difficult, because of the resonances. The pulsation frequencies interact and mutually depend on one another. Their amplitudes change on rather short time scales. We cannot unambiguously determine which are the independent parent frequencies, which detected signals are resonantly coupled child modes, and which are linear-combination frequencies. Also, if resonant phase-locking occurs, the detected frequency might be shifted compared to its unperturbed theoretical frequency.

We note that \cite{Moya2010MNRAS.405L..81M} determined the age of \oo\ using three frequencies published by Z99. However, as we mentioned above, two of these, detected around our $f_3$ might actually originate from a single mode that underwent amplitude and frequency variations during the observing run. Also, the frequency of the $f_3$ signal may be affected by resonance.

The only stable and relatively high-amplitude component of the pulsation of \oo\ is the dominant frequency, $f_1$. This can unambiguously be accepted as an independent pulsation mode. We only found two frequencies that appear to be independent from $f_1$. These are $f_7$ and $f_9$, however, both are of low amplitude.

The primary goal of the \most\ observations of \oo\ was to find as many independent pulsation modes as possible, in order to facilitate asteroseismic modelling, in particular, to determine the asteroseismic age of the system, independently from the work of \cite{Moya2010MNRAS.405L..81M}. Due to the mentioned problems, the asteroseismic age determination seems to be difficult to achieve at this point. As we mentioned in the introduction, we also collected rather extended high-resolution spectroscopic data on \oo\ during the 2008--2009 seasons. The analysis of these data will hopefully help to identify several independent pulsation modes, necessary to proceed with the asteroseismic age determination.

\begin{acknowledgements}

\'A.S. acknowledges support by the Belgian Federal Science Policy (project M0/33/029, PI: P.D.C.) and by the J\'anos Bolyai Research Scholarship of the Hungarian Academy of Sciences.
W.W. acknowledges support by the Austrian Science Fonds (FWF) P22691-N16.

\end{acknowledgements}



\bibliographystyle{aa}
\bibliography{sodor.bib}


\end{document}